\documentclass[9pt,twocolumn,twoside]{pnas-new}
\setboolean{displaywatermark}{false} % Set to false to remove the watermark
\fancypagestyle{firststyle}{}
\usepackage{upgreek}

\templatetype{pnasresearcharticle} % Choose template 
\title{Designing angle-independent structural colors using Monte Carlo
  simulations of multiple scattering}

\author[a]{Victoria Hwang}
\author[a]{Anna B. Stephenson} 
\author[b]{Solomon Barkley}
\author[a]{Soeren Brandt}
\author[a]{Ming Xiao}
\author[a,c,d]{Joanna Aizenberg}
\author[a,b,*]{Vinothan N. Manoharan}

\affil[a]{Harvard John A. Paulson School of Engineering and Applied Sciences, Harvard University, 29 Oxford Street, Cambridge, Massachusetts 02138, USA}
\affil[b]{Department of Physics, Harvard University, 17 Oxford Street, Cambridge, Massachusetts 02138, USA}
\affil[c]{Department of Chemistry and Chemical Biology, 12 Oxford Street, Cambridge, MA 02138, USA}
\affil[d]{Wyss Institute for Biologically Inspired Engineering, Harvard University, 60 Oxford Street, Cambridge, MA 02138, USA}

\leadauthor{Hwang} 

\correspondingauthor{\textsuperscript{*} E-mail: vnm@seas.harvard.edu}

\begin{abstract}
  {Disordered nanostructures with correlations on the scale of visible
    wavelengths can show angle-independent structural colors. These
    materials could replace dyes in some applications because the color
    is tunable and resists photobleaching. However, designing
    nanostructures with a prescribed color is difficult, especially when
    the application---cosmetics or displays, for example---requires
    specific component materials. A general approach to solving this
    constrained design problem is modeling and optimization: using a
    model that predicts the color of a given system, one optimizes the
    model parameters under constraints to achieve a target color. For
    this approach to work, the model must make accurate predictions,
    which is challenging because disordered nanostructures have multiple
    scattering. To address this challenge, we develop a Monte Carlo
    model that simulates multiple scattering of light in disordered
    arrangements of spherical particles or voids. The model produces
    quantitative agreement with measurements when we account for
    roughness on the surface of the film, particle polydispersity, and
    wavelength-dependent absorption in the components. Unlike discrete
    numerical simulations, our model is parameterized in terms of
    experimental variables, simplifying the connection between
    simulation and fabrication. To demonstrate this approach, we
    reproduce the color of the male mountain bluebird (\textit{Sialia
      currucoides}) in an experimental system, using prescribed
    components and a microstructure that is easy to fabricate. Finally,
    we use the model to find the limits of angle-independent structural
    colors for a given system. These results enable an engineering
    design approach to structural color for many different
    applications.}
\end{abstract}

\dates{This manuscript was compiled on \today}
\doi{\url{www.pnas.org/cgi/doi/10.1073/pnas.XXXXXXXXXX}}

\begin{document}

\maketitle
\thispagestyle{firststyle}
\ifthenelse{\boolean{shortarticle}}{\ifthenelse{\boolean{singlecolumn}}{\abscontentformatted}{\abscontent}}{}

% If your first paragraph (i.e. with the \dropcap) contains a list environment (quote, quotation, theorem, definition, enumerate, itemize...), the line after the list may have some extra indentation. If this is the case, add \parshape=0 to the end of the list environment.

\begin{figure*}[htbp]
\centering
\includegraphics{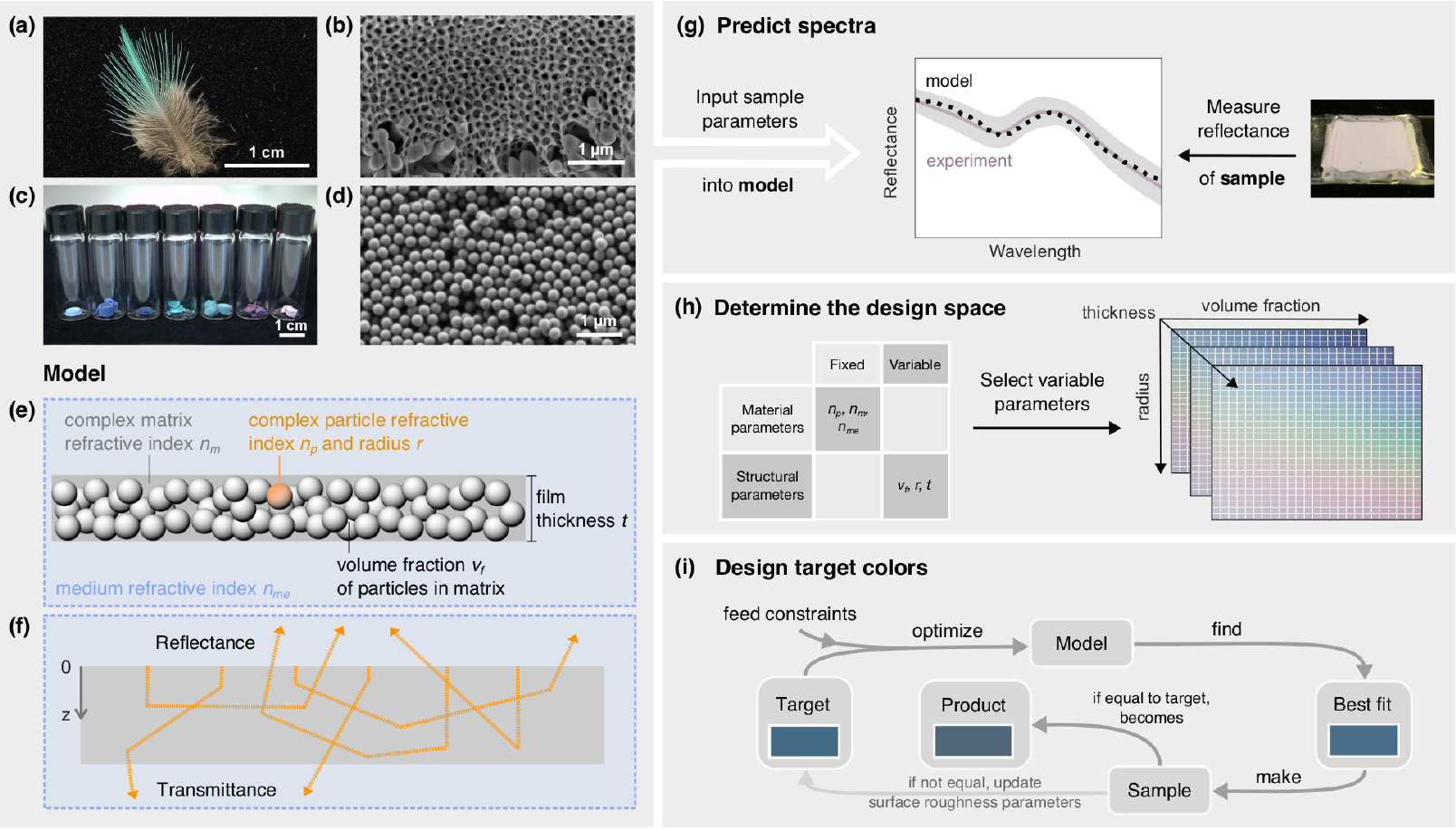}
\caption{Overview of design approach. (a) Photograph and (b) scanning
  electron micrograph (SEM) of features from a male Abyssinian roller
  (Specimen MCZ:Orn:63369. \textit{Coracias abyssinica}. Africa: Sudan:
  Blue Nile. El Garef. John C. Phillips; image credit: Museum of
  Comparative Zoology, Harvard University, \copyright President and
  Fellows of Harvard College). (c) Photographs of disordered packings of
  polystyrene particles, showing the structural colors that arise. The
  radii of the particles increase from left to right. (d) SEM of a
  disordered packing of 138-nm-radius polystyrene particles. (e)
  Schematic of the geometry used in our multiple-scattering model. The
  model is parameterized in terms of experimentally measurable
  quantities: the volume fraction, complex index of refraction, and
  radius of the spheres; the complex index of the matrix they are
  embedded in; the thickness of the film; and the index of refraction of
  the medium that lies between the viewer and the sample. (f) We
  calculate the reflectance spectrum by simulating thousands of photon
  trajectories, a few of which are shown schematically here. (g) The
  model can predict reflectance spectra that quantitatively agree with
  experimental measurements, as shown in the plot at center. Gray area
  is the uncertainty in the measurement. At right is a photograph of the
  measured sample. (h) The model can therefore be used to determine the
  design space---or all the possible colors---for specific constraints,
  such as a given type of particle and matrix material. Shown are
  examples of the colors that can be obtained for fixed material
  parameters and variable structural parameters. (i) Then, given a
  target color that is inside the design space, we use optimization to
  determine the experimental parameters needed to make that color,
  subject to constraints of our choice, as shown in this schematic.}
\label{fig:overview}
\end{figure*}

\dropcap{A}ngle-independent structural color occurs when light scatters
from a composite material with a correlation length on the scale of
visible wavelengths. Examples of such materials are the feathers of blue
and some green birds (Fig.~\ref{fig:overview}a), as well as disordered
packings of colloidal particles~\cite{forster_biomimetic_2010} with
radii around 100 to 150 nm (Fig.~\ref{fig:overview}b). The short-range
correlations between the pores in the bird feathers
(Fig.~\ref{fig:overview}c) and the particles in the colloidal sample
(Fig.~\ref{fig:overview}d) give rise to constructive interference of
backscattered light over a broad range of scattering wavevectors
$\mathbf{q}$~\cite{magkiriadou_absence_2014}. This broad range is
directly responsible for the weak angle-dependence of the colors, since
the scattering wavevector $\lvert\mathbf{q}\rvert=4\pi\sin\left(\theta/2
\right)/\lambda $ couples the scattering angle $\theta$ and the
wavelength $\lambda$. In comparison to the sharp Bragg peaks that occur
when light scatters from a material with long-range order, such as a
colloidal crystal, the reflection of a material with short-range order
peaks at a lower intensity. Nonetheless, short-range order can give rise
to vivid colors (Fig.~\ref{fig:overview}a). Furthermore, the weak
angle-dependence means that the color is almost indistinguishable from
that produced by an absorbing dye or pigment.

We can therefore envision replacing traditional dyes and pigments with
angle-independent structural colors in applications ranging from paints,
coatings, and cosmetics to electronic displays and sensors. There are
several advantages to doing so. Because structural colors do not require
strong absorption, they resist photobleaching and are less
photoreactive. Also, because structural colors arise from scattering,
different colors can be made from the same component materials by
changing the sizes of the pores or particles. This property enables a
new and powerful approach to formulation: first, the component materials
for a particular application are chosen to meet constraints such as low
toxicity or reactivity; second, the nanostructure is tuned to achieve
the target color.

Making this approach a reality requires a way to design materials with a
target color, starting from a prescribed set of component materials and,
frequently, a particular type of nanostructure. Many studies on
angle-independent color deal with a less constrained design problem:
given a color, determine the materials or nanostructure required to make
it. A common approach to address this less constrained problem is
biomimicry. Examples of biomimicry include fabricating structurally
colored materials containing melanin, an absorber found in bird
feathers~\cite{shawkey_interactions_2017}, to increase the
saturation~\cite{xiao_bio_inspired_2015, xiao_stimuli-responsive_2016,
  kohri_biomimetic_2015, wu_dopamine-melanin_2015,
  kawamura_structural_2017} or making nanostructures that mimic those
found in butterfly wings~\cite{zhang_inspiration_2015}. But biomimicry is
not a general approach to making structurally colored materials. In some
applications, a biomimetic system may not be compatible with the
constraints: for example, reflective displays might require
non-absorbing materials, so that the display does not heat when
illuminated. In other applications, the nanostructure might be too
difficult or expensive to fabricate.

Here we present a way to solve a more common and challenging design
problem: given the materials and a simple, easy-to-make
nanostructure---spherical inclusions in a matrix---determine what colors
can be made, and what structural parameters (particle size and volume
fraction, for example) are required to make a given color. Compared to
the problem of determining the materials and structure required to make
a given color, our problem is complicated by the potential absence of
solutions that meet the needs of the application. Determining whether a
solution exists requires a way to predict all possible colors for a
given set of constraints.

Tackling this design problem requires an accurate model for how light
interacts with disordered composite materials. Such a model must account
for both absorption and multiple scattering of light, which affects the
color saturation. Because structurally colored samples are composite
materials, each component has its own wavelength-dependent index of
refraction and absorption coefficient. Even if the components are
``transparent'' dielectrics, a small amount of absorption can change the
reflectance spectrum. As we shall show, these effects must be modeled
carefully to quantitatively predict color.

In what follows, we describe a Monte Carlo model of light transport in
disordered packings of spherical inclusions in a matrix phase, and we
show that it can make predictions that are in quantitative agreement
with experiment. We then use the model to determine the design space, or
all the possible colors that can be made, given the experimental
constraints. We demonstrate the design of target colors in two ways: in
the first, we choose target colors from the design space for specific
material systems. In the second, we target a given point in a perceptual
colorspace and use optimization to determine the experimental parameters
that produce this color. Overall, the approach that we demonstrate
(illustrated in Fig.~\ref{fig:overview}e--i) gives us the freedom to
precisely design and control angle-independent colors under constraints
of our choice, which opens possibilities that go well beyond biomimicry.

\section*{Model}
\label{sec:models}

\subsection*{Approach}
Light propagating through a composite material is scattered when the
refractive indices of the component materials differ and is absorbed
when either of the materials has a refractive index with a non-zero
imaginary component. The scattered waves can then interfere with one
another. Furthermore, depending on the refractive indices and
nanostructure, light might scatter repeatedly before exiting the
material. Thus, modeling structural color requires knowing the complex
refractive indices of the materials, the nanostructure, and the
detection geometry.

There are many approaches to modeling the relation between scattering
and color. The most venerable is radiative transfer theory, and in
particular the Kubelka-Munk theory~\cite{kubelka_beitrag_1931}, which
has been used extensively for predicting colors in mixtures of
paints~\cite{klein_industrial_2010, diebold_application_2014}. However,
radiative transfer theory does not in general capture interference
effects characteristic of structurally colored materials. Numerical
methods such as finite-difference-time-domain and finite-element
techniques~\cite{yin_amorphous_2012, dong_structural_2010,
  lo_structural_2014, cheng_structural_2015, galinski_scalable_2017,
  xiao_bioinspired_2017, chandler_structural_2017} do capture such
effects but are computationally intensive and difficult to use in design
because they are not parameterized in terms of experimental properties.
Approaches with a more natural parameterization include
single-scattering models based on effective-medium
approximations~\cite{magkiriadou_absence_2014,
  hwang_stephenson_effects_2019, maiwald_ewald_2018}. These models can
predict the wavelength of the reflection peak, but they do not account
for multiple scattering, which controls the color
saturation~\cite{hwang_stephenson_effects_2019}.

A more general approach is Monte Carlo simulation of photon
trajectories. In this approach, photon ``packets'' propagate through a
system, taking steps that are sampled from a step-size distribution and
scattering into directions sampled from a phase function
\cite{wang_mcml_1995, zolek_optimization_2006}. Monte Carlo methods have
been used to model multiple scattering in a variety of
systems~\cite{ding_influence_2016, dhawan_optical_2010,
  vinckenbosch_monte_2015, zhu_review_2013}, but they generally do not
account for constructive interference. Furthermore, achieving
quantitative agreement with experimental data requires a careful choice
of the step-size distribution and phase function, as we shall show.

\begin{figure*}[htbp]
\centering
\includegraphics{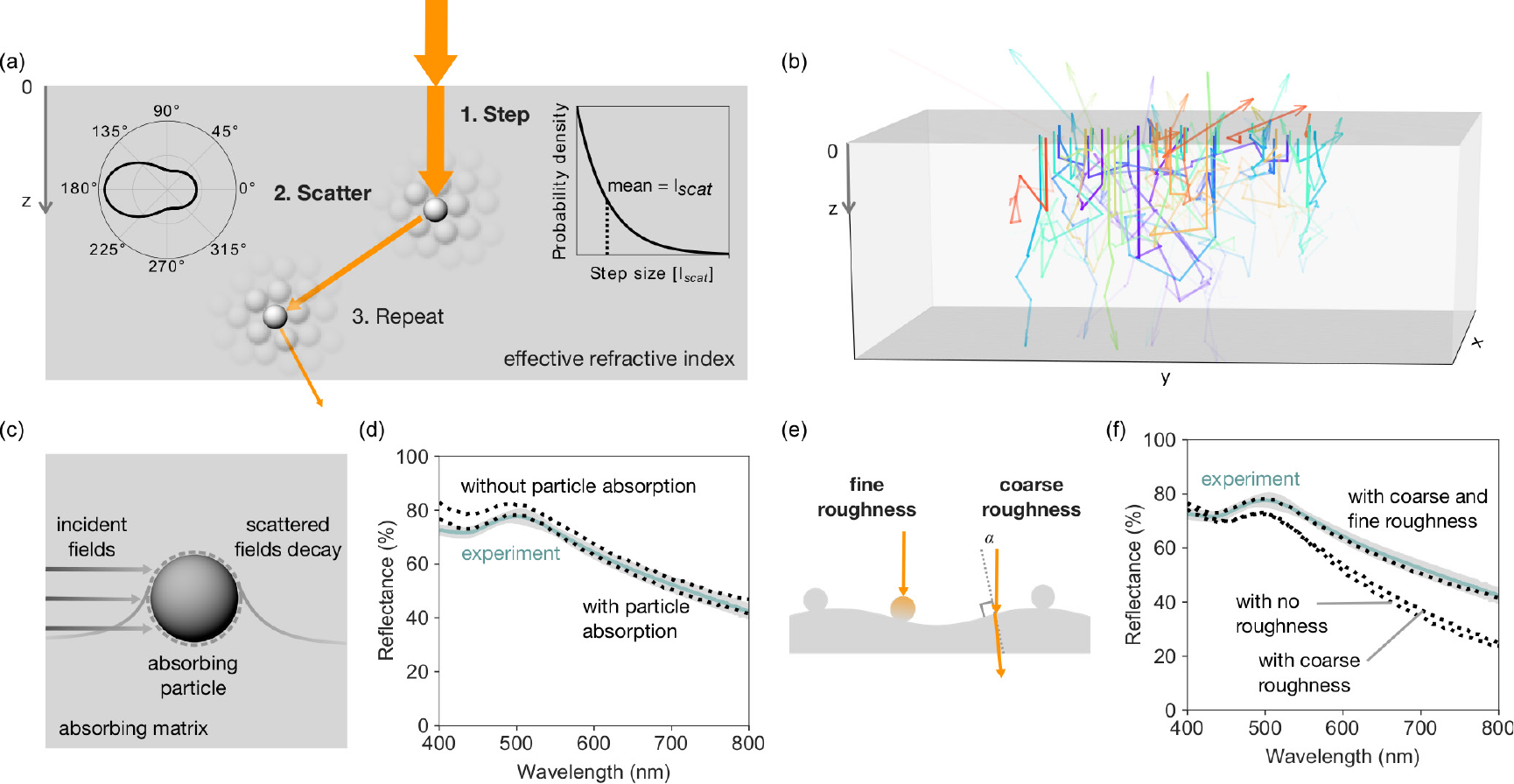}
\caption{Overview of Monte Carlo model for angle-independent structural
  color. (a) Cartoon of Monte Carlo method. We simulate the trajectories
  of photon ``packets'' scattering and propagating in an effective
  medium. Each packet first takes a step into the sample, where the step
  size is sampled from an exponential probability distribution whose
  mean is the scattering length. Part of the packet can be absorbed
  during this step, as illustrated by the decrease in the width of the
  orange arrow and as determined by the complex effective refractive
  index. Then the packet scatters into a new propagation direction,
  which is sampled from the phase function. Both the step-size
  distribution and phase function are calculated using the form and
  structure factors in the complex effective medium, and both depend on
  the wavelength. This process repeats until the packet exits the film.
  (b) Rendering of photon trajectories for a Monte Carlo simulation
  obeying the above rules. After simulating thousands of trajectories at
  different wavelengths, we calculate the reflectance spectrum by
  counting the fraction of packets that are reflected at each
  wavelength, as opposed to transmitted or absorbed. (c) Schematic
  showing how absorption is implemented in the model. In an absorbing
  system, the incident fields decay as they travel through the sample. The
  arrows depict the incident field and the curves represent the
  amplitude of the scattered fields. The dashed line is the surface of
  the particle, where we integrate the differential scattering
  cross-section. (d) Reflectance spectra calculated from the model,
  including and excluding the contribution of absorption, compared to
  experimental measurements of a 85-$\upmu$m film of 218 nm polystyrene
  spheres in air. Gray regions show the uncertainty on the measurement
  (see SI Appendix). (e) Diagram showing how roughness is modeled. The
  fine roughness parameter is the fraction of light that encounters
  roughness on the scale of a single particle upon incidence on the
  film, and is between 0 and 1. Coarse roughness corresponds to a
  tilted, though locally smooth, surface. The coarse roughness parameter
  is the root-mean-squared slope of the
  surface~\cite{van_ginneken_diffuse_1998}, and is 0 for a flat surface.
  While there is no upper bound, a large slope means that light cannot
  hit the sample; therefore, most systems have a coarse roughness
  between 0 and 1.1. (f) Reflectance spectra including and excluding the
  contribution of surface roughness, compared to experimental
  measurements of a 85-$\upmu$m film of 218 nm polystyrene spheres in
  air. Gray regions show the uncertainty on the measurement (see SI
  Appendix).}
\label{fig:model}
\end{figure*}

Our multiple scattering model is based on the Monte Carlo method, but we
use a phase function that accounts for constructive interference and
wavelength-dependent absorption, which we describe in more detail below.
With this phase function and the step-size distribution, we simulate the
random-walk trajectories of photon packets as they propagate through the
material, as shown in Fig.~\ref{fig:model}a. We consider the material to
be a film containing a disordered arrangement of spherical particles or
voids inside a matrix material. This film and the detector are embedded
in a medium, which we assume to be air in all the calculations that
follow. The film is parameterized in terms of both material and
structural quantities, including those shown in
Fig.~\ref{fig:overview}e. We assume that each packet is incident
normally on the film or at an angle determined by the experimental
setup, and we calculate the reflection spectrum by counting
trajectories, as shown in Fig.~\ref{fig:model}b.

\subsection*{Technique}
Our model accounts for constructive interference through the phase
function. There are two contributions to this function: the form factor,
which describes the angle-dependence of scattering from individual
particles and can be calculated from Mie theory; and the structure
factor, which describes the constructive interference of waves scattered
by different particles and can be calculated from liquid-state
theory~\cite{magkiriadou_absence_2014}. Although we account for
constructive interference within each trajectory through the structure
factor, we do not model constructive interference among different
trajectories.

We assume that the scattering occurs in an effective medium determined
by the average properties of the material (Fig.~\ref{fig:model}a). Our
effective-medium theory, which is described in more detail in SI
Appendix, is based on the Bruggeman
approximation~\cite{markel_introduction_2016}, which can account for
complex refractive indices, though not for near-field effects (see
Discussion).

To account for wavelength-dependent absorption, we use a modification of
Mie theory that accounts for absorption in both the particles and
matrix. In an absorbing system, the scattered fields are absorbed as
they propagate away from the scatterer, such that the differential
scattering cross-section of a particle decreases with
distance~\cite{bohren_absorption_2004,fu_mie_2001,mundy_mie_1974}. This
consideration applies not only to systems with absorption in the matrix,
but also to those with only absorbing particles, because in both cases
the imaginary index (and hence the absorption coefficient) of the
effective medium is non-zero. Therefore, when the particle or matrix has
a complex refractive index, we obtain the total scattering cross-section
by integrating the differential scattering cross-section at the surface
of the scatterer~\cite{sudiarta_mie_scattering_2001,
  frisvad_computing_2007}. We then account for absorption of the photon
packets traveling through the effective medium with an exponential decay
function based on the Beer-Lambert law. Lastly, we correct for the
variation in the amplitude of the incident field as a function of
position on the sphere~\cite{sudiarta_mie_scattering_2001}
(Fig.~\ref{fig:model}c). For more details on the model, see SI Appendix. 

Modeling absorption leads to better agreement between the predicted and
measured reflection spectrum (Fig.~\ref{fig:model}d). The small amount
of absorption in polystyrene particles, for example, changes the
predicted reflectance spectrum from that of a sample without absorption,
especially at short wavelengths.

We also account for surface scattering, which can arise from the
roughness inherent to most experimental samples. We model this roughness
at two different scales: coarse and fine. Coarse roughness is large
compared to the wavelength, such that incident light encounters a
locally smooth surface that is angled with respect to the incident
direction. We model coarse roughness by accounting for the refraction of
light when it encounters the boundary of the film
(Fig.~\ref{fig:model}e). Fine roughness arises from wavelength-scale
features such as particles protruding from the surface. To model fine
roughness, we sample the initial step size of a trajectory from the
scattering properties of a single nanoparticle, ignoring the
contribution of the structure factor.

For many of the samples we examine, such as those dispersed in a liquid,
we cannot easily measure the roughness parameters. Indeed, as we note in
Discussion, the roughness parameters can be viewed more generally as
correcting for the failure of the effective-medium approximation at the
boundary of the sample. Therefore we determine these parameters by
fitting them to measurements. When we do this, we find that including
coarse and fine roughness brings the model into quantitative agreement
with experiment (Fig.~\ref{fig:model}f). Although the parameters are
fitted, they are constant with wavelength. Therefore the agreement
between the fitted model and the data as a function of wavelength shows
that our model for roughness captures a physical effect of the sample
boundary.

In the validation experiments that follow, we do not fit each
measurement individually; instead, because we expect the roughness
values to largely depend on the sample assembly method, we fit the
values to all samples fabricated with the same technique.

\section*{Results}
\label{sec:results}

\subsection*{Model validation}

\begin{figure*}[tbhp] 
\centering
\includegraphics{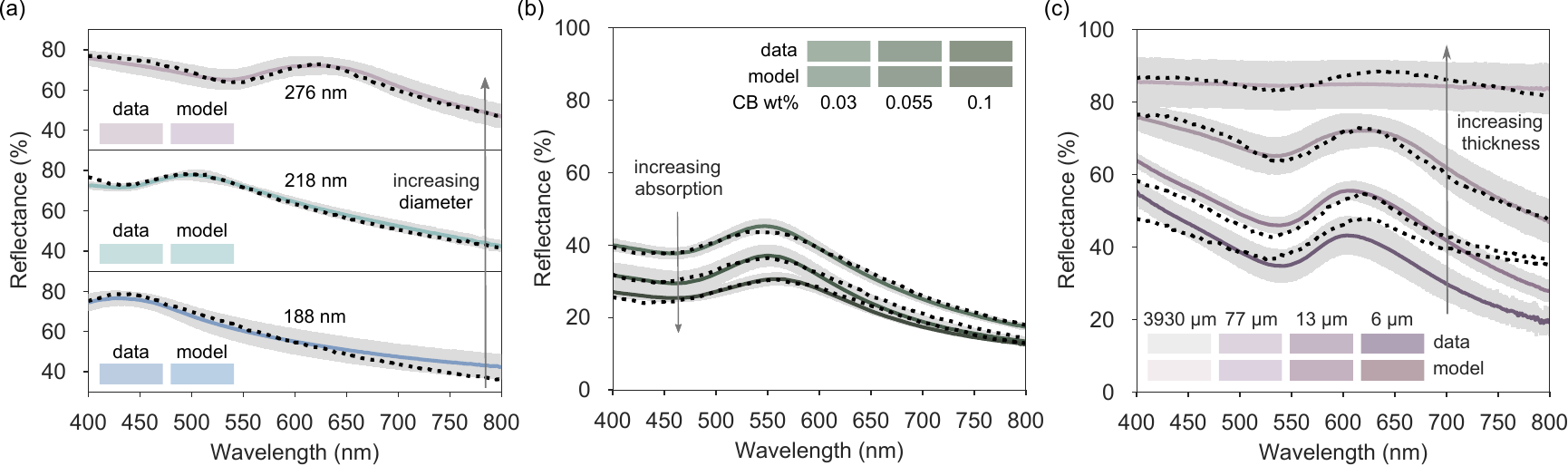}
\caption{Experiments validate the Monte Carlo model. Measured (solid
  lines) and predicted (dotted lines) reflectance spectra for disordered
  packings of polystyrene particles as a function of (a) particle
  radius, (b) absorption, and (c) film thickness. Gray regions show
  uncertainties on the measurements (see SI Appendix). Insets are color
  swatches calculated from the experimental and predicted spectra using
  the CIELAB colorspace. The model parameters are as follows: (a)
  Polystyrene spheres with radii of 94, 109, 138 nm in a matrix of air,
  with volume fractions of 0.52, 0.52, 0.56, and thicknesses of 119, 85,
  and 77 $\upmu$m. The fine roughnesses are 0.5, and the coarse
  roughnesses are 0.9 for all samples. (b) 101-nm-radius polystyrene
  particles in water with carbon black at concentrations of 0.03, 0.055,
  and 0.1\% by weight in water. The particle volume fractions are 0.415,
  0.406, and 0.386, and the thicknesses are 96, 71, and 84 $\upmu$m. The
  fine roughness is 0.28 and the coarse roughness is 0.2 for all three
  samples. (c) Films of 138-nm-radius polystyrene particles at
  thicknesses of 3930, 77, 13, and 6 $\upmu$m. The corresponding volume
  fractions are 0.5, 0.56, 0.58, 0.58. The fine roughnesses are 1 for
  the 3930-$\upmu$m film and 0.5 for all others, and the coarse
  roughnesses are 0.9. The thickness of the 3930-$\upmu$m film is chosen
  to be much larger than the maximum transport length, 47 $\upmu$m, to
  ensure strong multiple scattering. The thickness of the 6-$\upmu$m
  film is chosen to be smaller than the minimum transport length, 8
  $\upmu$m, to minimize multiple scattering.}
\label{fig:validation_results}
\end{figure*}

We validate the model by comparing the predicted and measured
reflectance spectra for samples with different physical parameters (see
SI Appendix for a description of the sample fabrication and
characterization). In each of the simulations, we take the average of
20,000 trajectories at each wavelength. For such a large number of
trajectories, the Monte Carlo uncertainty in the predicted reflection
spectrum is much smaller (standard deviation 0.4\%) than the uncertainty
of the experimental spectrum, which is determined by taking measurements
from different parts of the same sample.

We first examine the effect of particle radius. We calculate
reflectance spectra for packings of polystyrene particles in air for
three different polystyrene radii: 94, 109, and 138 nm. As shown in
Fig.~\ref{fig:validation_results}a, the model accurately captures the
redshift of the reflectance peak with increasing particle size while
also reproducing a rise in scattering with toward small wavelengths.
The predicted spectra quantitatively match the data in both the location
of the reflectance peak and the reflectance magnitude across the visible
range with only small deviations. The model also captures the broadening
and averaging of the peak when two particle radii are mixed, which
validates our implementation of polydispersity (Fig.~S1a). As a result,
the colors predicted by the model visually match the color renderings
calculated from the measured reflectance.

Having shown previously that a small amount of absorption can alter the
reflection spectrum (Fig.~\ref{fig:model}d), we must now further confirm
that our model accurately captures the effects of absorption in
experimental samples. We make concentrated samples of polystyrene
spheres in water, and we tune the amount of absorption by adding varying
amounts of carbon black. To model these samples, we assume a matrix with
a real refractive index of water and an imaginary index corresponding to
the concentration of carbon black (see SI Appendix). Thus we neglect any
scattering from the carbon black particles, which is a reasonable
approximation, given that the carbon black particles are approximately
10 nm, much smaller than the wavelength. We again find that the model
accurately predicts the reflectance and color of samples with varying
amounts of absorption (Fig.~\ref{fig:validation_results}b).

In addition, we explore the validity of the model over a range of film
thicknesses. In the thickest sample, the thickness is much larger than
its transport length, which is the length scale at which the direction
of light is randomized. In the thinnest, the thickness is smaller than
its transport length at all wavelengths. The model agrees with
experiment when the thickness is large, but starts to deviate from
experimental data in thin samples and at large wavelengths
(Fig.~\ref{fig:validation_results}c). The discrepancy likely arises
because for very thin samples, the distinction we make in our model
between surface scattering and bulk scattering starts to break down.
However, most structurally colored samples are not as thin as the
thinnest sample we show here. Furthermore, the predicted colors in all
samples are similar to those of the experimental samples, despite the
deviations in the predicted reflection spectrum for thin samples.

In SI Appendix, we further validate the model on bidisperse
samples (Fig.~S1a) and samples with varying volume fraction (Fig.~S1b).
For the volume-fraction experiments, we prepare samples of polystyrene
spheres in water, in which case the volume fraction can be varied by
changing the particle concentration.

\begin{figure*}[htbp] 
\centering
\includegraphics{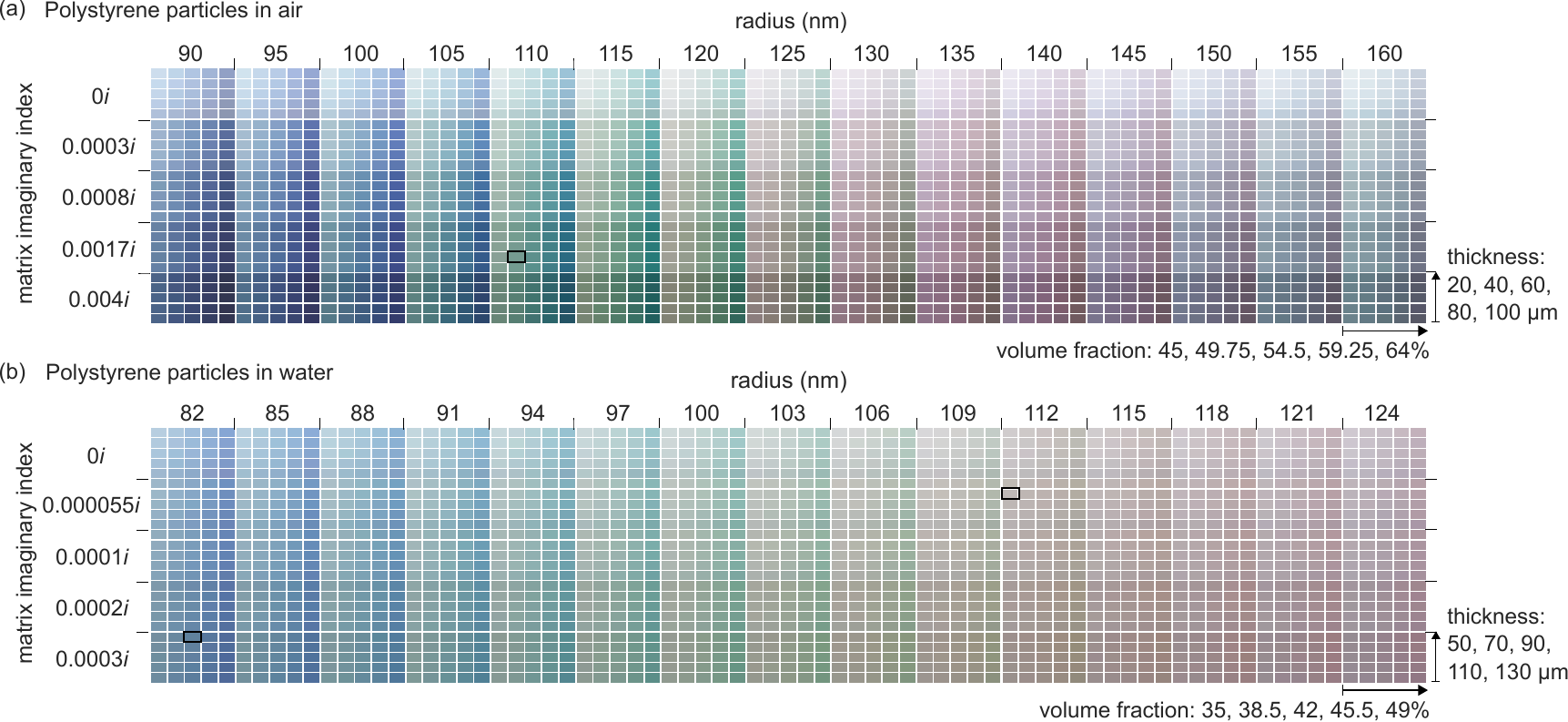}
\caption{The range of designable colors depends on the chosen materials.
  Each of the two grids is a representation of the color gamut in a
  four-dimensional parameter space spanned by the particle radius,
  matrix imaginary index (a proxy for carbon black concentration),
  particle volume fraction, and sample thickness. Each rectangle in the
  grid is a color swatch calculated from a reflection spectrum predicted
  by the Monte Carlo model. The rectangles are organized into 5$\times$5
  subgrids. Particle radius increases with subgrid from left to right,
  and matrix imaginary index increases with subgrid from top to bottom.
  Within each subgrid, volume fraction increases from left to right and
  thickness from bottom to top. (a) Color gamut for polystyrene
  particles in a matrix of air. We use the same roughness parameters as
  for the film of 276-nm particles in
  Fig.~\ref{fig:validation_results}a: coarse roughness of 0.9 and fine
  roughness of 0.5. (b) Color gamut for polystyrene particles in a
  matrix of water. We use the same roughness parameters as for the
  polystyrene-in-water films in Fig.~\ref{fig:validation_results}b:
  coarse roughness of 0.2 and fine roughness of 0.28. In both gamuts the
  imaginary refractive index for polystyrene is fixed at $2\times
  10^{-5}i$ and the polydispersity index of the polystyrene particles is
  0.03.}
\label{fig:gamuts}
\end{figure*}

\subsection*{Finding the limits of the design space}
With a validated model, we can calculate the limits of the design
space---that is, the range of structural colors that can be made for a
given set of materials or other constraints. As an example, we calculate
a gamut for packings of polystyrene particles in air with added carbon
black, with varying particle radius, volume fraction, sample thickness,
and carbon black concentration (Fig.~\ref{fig:gamuts}a).

To describe how the colors change as a function of these parameters, we
use terminology from color science: hue, chroma (or perceived
vividness), and luminance. Each of these can be calculated by
transforming the computed reflectance spectra used to generate
Fig.~\ref{fig:gamuts}a to the CIELUV perceptual colorspace. We find that
small particle radii give rise to colors in the blue and green, as
expected, but red hues remain inaccessible, in agreement with the
results of Schertel and coworkers~\cite{schertel_structural_2019}. We
also find that increasing the volume fraction can significantly increase
the chroma and blue-shift the hue while decreasing the luminance.
Increasing the thickness does not affect the hue. Instead, it slightly
increases the chroma and luminance at small imaginary indices but not at
the largest imaginary indices, where the absorption length becomes
comparable to or smaller than the sample
thickness~\cite{hwang_stephenson_effects_2019}. When we replace the air
matrix with water, we find that increasing the radii leads to a range of
browns instead of pinks and purples (Fig.~\ref{fig:gamuts}b), because
the lower index contrast between polystyrene and water leads to flatter
and broader reflectance peaks. Increasing the volume fraction
blue-shifts the hue and increases the chroma. The thickness does not
affect the hue or chroma, but only increases the luminance when the
absorption is low, as in the polystyrene-in-air system. In both systems,
increasing absorption only decreases the luminance and does not change
the hue or chroma.

\begin{figure}[htbp]
\centering
\includegraphics{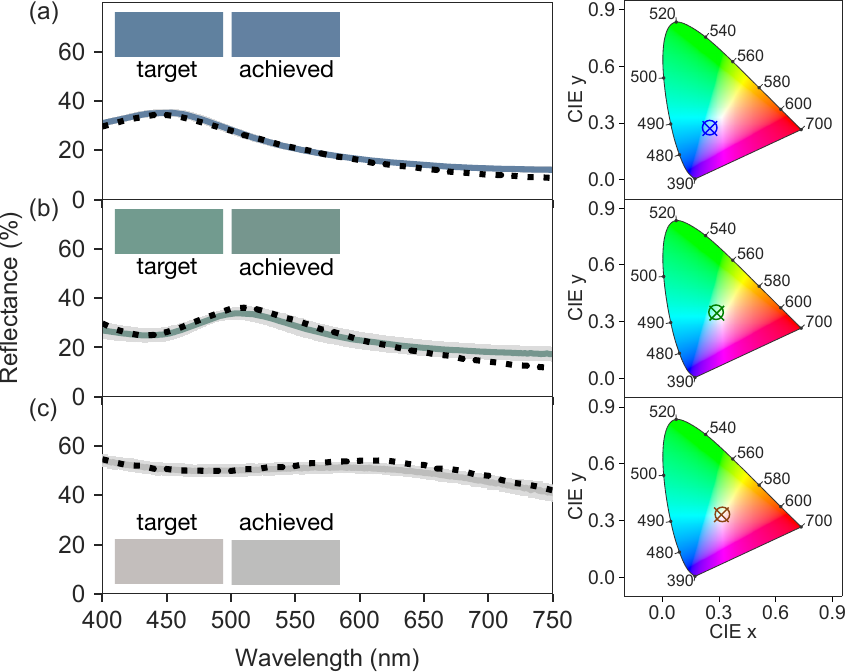}
\caption{Designing colors from the gamut. Each plot shows the
  reflectance spectra of the target color (dotted line) and of the color
  that is achieved (solid line) in a sample made using the model
  parameters for the target. To the right of each plot is a colormap
  showing the CIE chromaticity coordinates of the target (circles) and achieved
  (crosses) colors. The target colors are chosen from the color gamuts
  in Fig.~\ref{fig:gamuts}. The parameters are as follows. (a) Blue
  target: radius 82 nm, volume fraction 0.42, thickness 130 $\upmu$m,
  and matrix imaginary index 0.0003$i$, corresponding to 0.08\% by
  weight of carbon black. (b) Green target: radius 110 nm, volume
  fraction 0.4975, thickness 40 $\upmu$m, and matrix imaginary index
  0.0017$i$, corresponding to a carbon black concentration of 0.42\% by
  weight. (c) Brown target: radius 112 nm, volume fraction 0.35,
  thickness 110 $\upmu$m, and matrix imaginary index 0.000055$i$,
  corresponding to 0.016\% by weight of carbon black. The uncertainties
  in the achieved spectra are shown in gray and represent two standard
  deviations about the mean of measurements from 11 (blue spectrum), 19
  (green) and 12 (brown) locations on the sample.}
\label{fig:particle_design}
\end{figure}

To demonstrate the predictive power of the model, we make three colors
from these gamuts (outlined swatches in Fig.~\ref{fig:gamuts}). The
colors are chosen from across the visible spectrum. We make a green
sample with polystyrene particles in air, and a blue and a light brown
sample with polystyrene particles in water. We make samples with
parameters as close as possible to the values used in the simulations,
and we find that the target and the achieved colors agree well, with
some small deviations at large wavelengths
(Fig.~\ref{fig:particle_design}).

\subsection*{Finding the parameters to design a target color}

In addition to targeting colors in the gamut, we can also target a
particular color in a colorspace. We use the perceptual colorspace
defined by the CIELAB coordinates~\cite{cieluv2} because applications
such as cosmetics or coatings are aimed at the human eye. Using such an
approach increases the number of available designs because we can
exploit the eye's insensitivity to variations in certain parts of the
spectrum.

To implement this approach, we choose a target color in CIELAB
coordinates, then use Bayesian
optimization~\cite{fernando_bayesianoptimization_2019} to find the model
parameters that minimize the sum of squared differences between the
target CIELAB coordinates and those corresponding to the modeled
reflection spectrum. We call the optimal solution the ``best fit'' to
the target.

\begin{figure*}[htbp]
\centering
\includegraphics{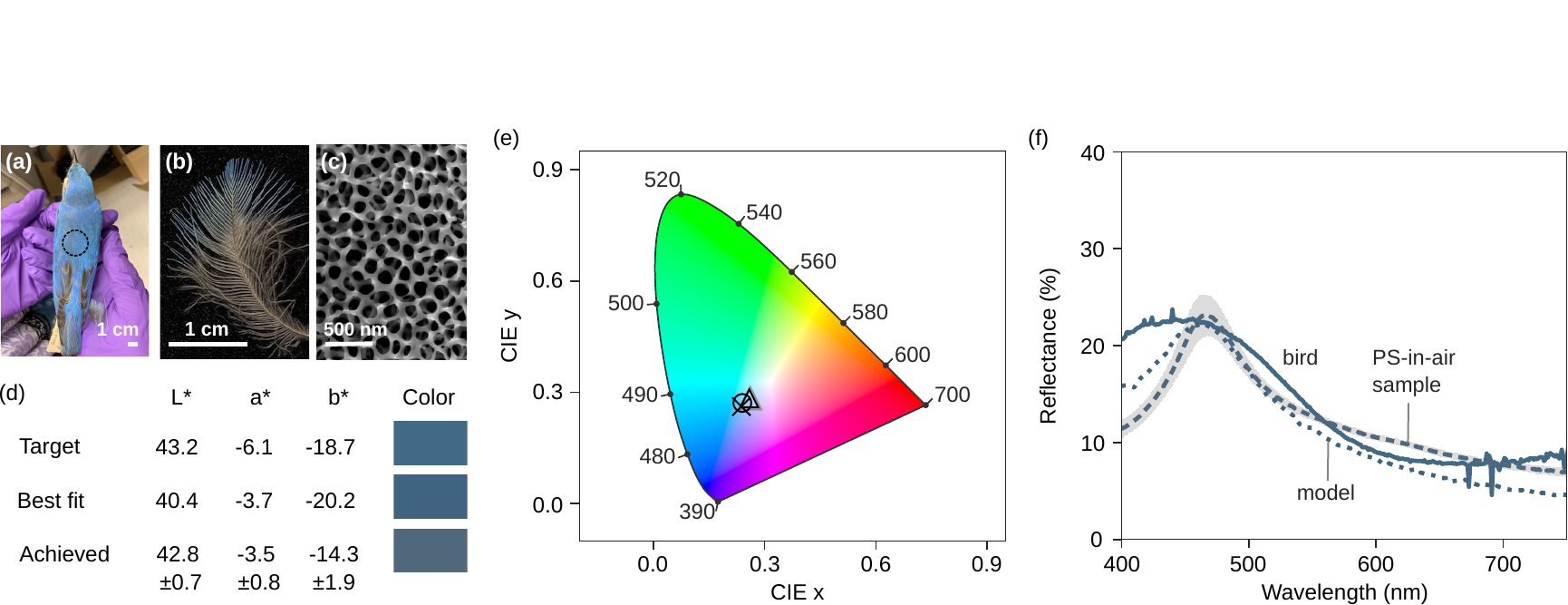}
\caption{Targeting a specific color in the colorspace. (a) Photograph of
  male mountain bluebird (Specimen MCZ:Orn:190556. \textit{Sialia
    currucoides}. North America: United States: Montana: Meagher.
  Martinsdale. Robert S. Willians). Circle
  denotes area of reflectance measurement for the target color. (b)
  Photograph of a feather from the back of the bird. (c) SEM of a
  cross-section of the feather's internal structure, obtained after
  focused-ion beam milling. Image credits for (a-c): Museum of
  Comparative Zoology, Harvard University, \copyright President and
  Fellows of Harvard College. (d) CIELAB values and color renderings of
  the target, best-fit, and achieved colors. The parameters of the
  best-fit solution that satisfies the constraints are as follows:
  101-nm-radius polystyrene spheres at a volume fraction of 0.54 in an
  air matrix, 50 $\upmu$m film thickness, 0.9\% by weight of carbon
  black, fine roughness of 0.5, and coarse roughness of 0.9. (e) CIE
  chromaticity color map comparing the target (circle), best-fit
  (cross), and achieved colors (triangle, uncertainty shown in gray).
  (f) Reflectance spectra of bird feather (solid line), model fit
  (dotted line), and polystyrene-in-air sample (dashed line). Gray
  regions for the achieved reflectance represent two standard deviations
  about the mean of measurements at 3 locations on the sample. Note that
  we do not try to match the model and target spectra; instead, the
  optimization is performed in the $L*, a*, b*$
  space.}
\label{fig:mountain_bluebird_design}
\end{figure*}

We choose the color of the mountain bluebird as our target
(Fig.~\ref{fig:mountain_bluebird_design}a), because the feathers show an
angle-independent structural blue
(Fig.~\ref{fig:mountain_bluebird_design}b). This color arises from a
porous internal structure (Fig.~\ref{fig:mountain_bluebird_design}c)
that likely evolved to meet constraints other than color, including
(perhaps) minimizing weight and maximizing insulating ability.

We impose a different set of constraints. Because the bluebird's
``inverse'' structure of pores inside a solid matrix is not as easy to
fabricate as a ``direct'' structure of solid spheres in air or water, we
design the color using a direct structure instead. Furthermore, we
constrain the materials to those we have on hand: polystyrene spheres
and a matrix of either air or water. We use Bayesian optimization to
determine the optimal particle radius, volume fraction, film thickness,
and concentration of carbon black. To ensure that the optimal values can
be experimentally achieved, we set ranges for these
parameters: the particle radius is 74, 101, 110, 112.5, or 138 nm; the
thickness is between 20 and 150 $\upmu$m; and the range for the matrix
imaginary index is between 0 and 0.005$i$. We use the same values of
roughness as in the samples in Fig.~\ref{fig:validation_results}a: fine
roughness of 0.5 and coarse roughness of 0.9.

When we minimize the difference between the target color and the color
obtained from the model, we find a good match in CIELAB space
(Fig.~\ref{fig:mountain_bluebird_design}d, e). Note that we match the
color and not the reflectance spectrum, because matching the spectrum
may not be possible for the given materials. Indeed, the spectrum of the
best-fit solution has a narrower peak than that of the target, with the
target having a larger reflectance at wavelengths less than 450 nm
(Fig.~\ref{fig:mountain_bluebird_design}f). Because the eye is
insensitive to such short wavelengths, the best-fit solution need not
duplicate this feature to match the color in the CIELAB space. We find
that the best-fit solution has a CIE76 color difference of 3.9 from the
target color, which is close to the just-noticeable difference (JND) of
2.3~\cite{sharma_digital_2002}.

We make a film with parameters as close as possible to those of the best
fit. We find that both the resulting spectrum and color are close to
those of the best fit, as shown in
Fig.~\ref{fig:mountain_bluebird_design}d-f. The CIE76 color difference
between the achieved and target colors is 5.1, larger than the
difference between target and best-fit, but still less than twice the
JND. The difference between the best-fit and achieved colors may come
from the values of the roughness parameters we use in the model.
Although we use the same preparation technique and thus the same fine
and coarse roughness values as for the polystyrene films from
Fig.~\ref{fig:validation_results}a, the true roughness values of the
sample might differ from these values. Nonetheless, the agreement
between the achieved and target colors shows that one can design the
color of the feather without mimicking its structure, instead using a
system that satisfies a different set of constraints. 

\section*{Discussion}

Having shown that our model can be used to design colors, we now examine
its limitations and why it works despite these limitations. The central
approximation is that of the effective medium: we assume that between
scattering events, light propagates through a homogeneous medium with an
effective refractive index. This effective index, which we calculate
using the Bruggeman weighted average~\cite{markel_introduction_2016},
underlies all of the calculations. For example, the index difference
that determines the phase function is the difference between the
effective index and the index of the spheres. An alternative
effective-index theory, called the energy coherent potential
approximation (ECPA), includes corrections for near-field effects in a
dense packing of spherical scatterers~\cite{busch_transport_1996,
  *busch_transport_1995}. Schertel and coworkers used the ECPA in
concert with the diffusion approximation to predict structural
colors~\cite{schertel_structural_2019}.

Their model, like ours, can predict reflectance spectra, but it is not
suitable for our purposes for two reasons. First, the diffusion
approximation is valid only when light scatters many times before it
exits the sample, whereas in many structurally colored samples, the
thickness is chosen to minimize the amount of multiple
scattering~\cite{hwang_stephenson_effects_2019}. Second, the ECPA is
valid only for real dielectric permittivities, and therefore does not
account for absorption. Schertel and coworkers compensate for this
limitation by approximating absorption as a cutoff of the sample
thickness. Our approach uses the Bruggeman effective medium
approximation, which has the disadvantage that it does not account for
near-field effects, but the advantage that it can account for complex
refractive indices. Therefore it can directly handle absorption and its
dependence on wavelength.

To understand why our approach correctly predicts spectra despite the
absence of near-field corrections, we calculate the scattering strength
of polystyrene particles in air. The scattering strength is the ratio of
the wavelength to the transport length, where the transport length is
calculated using both the form and structure factors and the Bruggeman
effective index. We calculate the scattering strength as a function of
the ratio of radius to wavelength. We find that the peaks in scattering
strength match experimental measurements by Aubry and
colleagues~\cite{aubry_resonant_2017} up to a radius-to-wavelength
ratio of roughly 0.5 (Fig.~S2), which covers the range used in our
study. Near-field effects can be neglected in this range because the
experimental transport length is at least four times larger than the
wavelength.

Because our model does not account for near-field effects, we do not
expect it to work at high scattering strengths, or when the transport
length is comparable to the wavelength. Furthermore, the discrepancies
at small and large thicknesses (Fig.~\ref{fig:validation_results}c)
suggest that our model works best in the regime of weak multiple
scattering, where the film thickness is on a similar order of magnitude
as the transport length. When multiple scattering is stronger,
interference between multiply scattered photon trajectories, which our
model does not account for, might become important.

However, these situations may not be relevant to structurally colored
materials. Color saturation requires that multiple scattering be weak
and the transport length be comparable to the thickness of the material.
This is the fundamental reason why the model works so well. Furthermore,
we take advantage of the limited color capacity of the human visual
system---especially at long wavelengths---because our eyes detect colors
based on three receptors rather than on full reflectance spectra.
Therefore, our model can be used to design perceptual colors even when
the reflection spectrum cannot be matched to the target spectrum, as we
have shown.

We have also shown that it is necessary to model surface roughness to
achieve quantitative agreement between model and experiment. When we
introduced the roughness parameters, we argued that they account for the
topography of the samples. But more generally, the roughness parameters
account for the breakdown of the effective-medium approximation at the
surface of the sample. The breakdown occurs not only because of
topography, but because Mie theory does not accurately describe the
initial interaction of light with the film. Mie theory is derived for a
particle embedded in the same (effective) medium on all sides, whereas
particles at the surface have other particles on one side and a
homogeneous material on the other. Furthermore, the effective-medium
approximation we use includes the effects of the structure factor, which
is not well defined at the boundary of the sample. The fine and coarse
roughness parameters compensate for all of these effects, and therefore
topographical measurement techniques such as atomic force microscopy may
not give the appropriate values for these parameters.

Nevertheless, the model is still predictive even though the roughness
parameters must be fitted to experimental data. Indeed, as we have
shown, the parameters need not be fitted to measurements for each
individual sample; instead, one can use the same values for all samples
that are made with the same assembly technique. To improve the
predictive accuracy, one can use an iterative design approach: first,
make an initial guess for the roughness and find the model
parameters that best fit a target color; second, make the sample using
the best-fit parameters and fit the model to the data to improve the
estimates of the roughness parameters; third, use the improved estimates
to find parameters that give a better fit of the model to the target
(Fig.~\ref{fig:overview}d).

\begin{figure}[htbp] 
\centering
\includegraphics{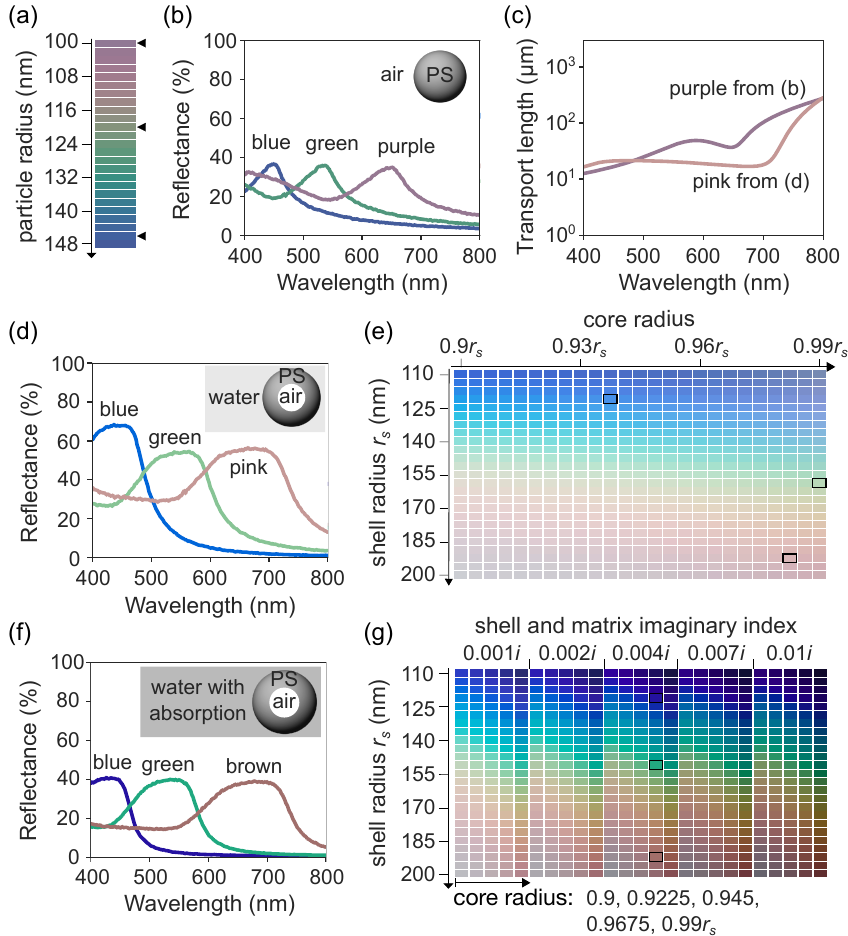}
\caption{The model allows us to explore the limits of colors that can be
  achieved with different configurations of a set of materials. All
  plots assume an imaginary refractive index for polystyrene of $2\times
  10^{-5}i$, volume fraction of 0.64, thickness of 20 $\upmu$m, coarse
  roughness of 0.9, and fine roughness of 0.5. (a) Color gamut for
  polystyrene (PS) particles in a matrix of air as a function of
  particle radius. (b) Reflectance spectra for colors denoted by the
  black arrows in (a). (c) Transport length as a function of wavelength
  for the purple system in (b) and the pink system in (d). (d)
  Reflectance spectra for three colors chosen from the gamut for
  particles with air cores and PS shells in a matrix of water. (e) Color
  gamut for the core-shell system as a function of core and shell
  radius. Circles denote the samples whose reflectance spectra are shown
  in (d). (f) Reflectance spectra for three colors from the gamut of a
  core-shell system with absorption added to the matrix. (g) Color gamut
  for core-shell system as a function of core radius, shell radius, and
  matrix imaginary index. Circles denote the samples whose reflectance
  spectra are shown in (f).}
\label{fig:loosening_constraints}
\end{figure}

The power of our model lies in providing a physical understanding of how
the experimental parameters change the color. This insight enables a
rational design approach for the nanostructure. Consider a case when the
materials are prescribed---for example, polystyrene in air or
water---but the structure can be varied---for example, by making
composite particles. This situation arises in many applications: the
constituent materials must meet certain requirements (regulatory or
other), but the spatial arrangement of these materials may be
unconstrained. Because there are infinite possible arrangements that
differ from solid spherical particles in a matrix, finding the optimal
arrangement for a target color is a very difficult design problem. We
can, however, use the physical intuition provided by the model to choose
a nanostructure that produces a particular color.

As an example, we consider making colors that are outside the gamut of a
system of polystyrene particles in air, yet use the same materials.
Solid polystyrene spheres in air tend to have low saturation or chroma,
particularly in the red, as shown in the gamut of
Fig.~\ref{fig:loosening_constraints}a. The low saturation comes from
scattering at short wavelengths, as shown in the purple spectrum in
Fig.~\ref{fig:loosening_constraints}b. The short-wavelength scattering
comes from the large scattering cross-section of polystyrene particles
in the blue~\cite{hwang_stephenson_effects_2019}. The model shows that
this large cross-section gives rise to multiple scattering. The
propensity for multiple scattering can be described by the transport
length, which is small at short wavelengths
(Fig.~\ref{fig:loosening_constraints}c).

To decrease this scattering, we design an alternative arrangement of the
materials. First, we invert the particles into air cores with
polystyrene shells~\cite{magkiriadou_absence_2014} to reduce the
scattering cross-section in the blue. Second, we place the core-shell
particles in a matrix of water to decrease the index contrast between
the shell and the matrix (Fig.~\ref{fig:loosening_constraints}d).
Because the resulting colors are still desaturated
(Fig.~\ref{fig:loosening_constraints}e), we suppress multiple scattering
by adding an absorber to the water
(Fig.~\ref{fig:loosening_constraints}f). We use the model to determine
what absorber concentrations lead to optimal saturation. The resulting
gamut shows colors and saturations that are different from those of
polystyrene particles in air---in particular, we now see orange and
brown hues that arise due to the decreased scattering at short
wavelengths (Fig.~\ref{fig:loosening_constraints}g).

From this example, we see that loosening the restrictions on the
arrangement of the materials increases the size of the design space but
also makes it possible to access new colors. The physical intuition
provided by the model is critical for exploring this larger design
space. 

\section*{Conclusion}
\label{sec:conclusion}

Engineering materials with prescribed structural colors requires a way
to predict the color from the nanostructure. Doing so efficiently
requires accurate predictions, so as to minimize iteration between
experiment and simulation. We have demonstrated a model of multiple
scattering in disordered packings of spheres that produces accurate
predictions. The model is parameterized in terms of experimental
quantities such as the volume fraction and material optical properties.
As such, it can be used to design structurally colored materials that
meet specific constraints, making it particularly useful for
applications in which only certain materials can be used. The predictive
power is also important for applications such as paints and coatings,
where the color might change substantially as the suspending liquid
dries and the refractive-index contrast increases.

Compared to finite-difference time-domain and similar methods, our model
addresses a smaller range of nanostructures---those involving disordered
arrangements of spheres---but has two principal advantages: it is
parameterized in terms of experimental quantities, and the results can
be interpreted in terms of collective and resonant scattering effects.
The interpretability of the results can be used to rationally design
variant nanostructures with wider color gamuts, as we have shown.
Furthermore, the restriction to spherical pores or voids may not be a
significant limitation. A feature of angle-independent structural color
is that it does not require a complex nanostructure or fabrication
scheme; instead, it can be produced simply by rapidly consolidating
inexpensive spherical nanoparticles. We anticipate that many
applications will take advantage of this feature.

We have validated the model in the regime of weak multiple scattering,
that is, when the transport length is several times the wavelength, and
the sample thickness is on the order of the transport length. Arguably
most applications of structural color lie within this regime, including
paints, coatings, and sensors. Materials that strongly multiply scatter
light look white, whereas materials that have little multiple scattering
look translucent. Other models---for example, the diffusion
approximation in the case of strong multiple scattering and
single-scattering theory in the case of very weak multiple scattering---
may yield better predictions in these two regimes.

Our model can be extended to more complex geometries and other
illumination conditions. Here we have focused on films of spheres in a
matrix, but the boundary conditions can be changed to model so-called
``photonic balls''~\cite{yi2003generation, moon2004electrospray,
  vogel2015color} or even films of photonic balls. Also, because the
model accounts for dispersion and wavelength-dependent absorption, it
can be used to design materials with infrared or ultraviolet reflection
peaks, so long as the transport length remains large compared to the
wavelength.

Finally, the model can be used to predict the angle-dependence of the
color. Although the term ``angle-independent'' is used to describe the
color, in reality there is a weak variation of color with the angle
between the source and detector. This variation arises because 
constructive interference condition is not completely independent of the
scattering wavevector~\cite{magkiriadou_absence_2014}. In the results
shown above, we have simulated the diffuse reflectance spectrum, which
can be used to determine how the sample would look in ambient light. But
one can also simulate the reflection spectrum as a function of the
incident and detected angles. This approach could help determine the
fundamental physical limits of angle-dependence, chroma, and hue---as
well as the tradeoffs between them---in structurally colored materials.

\matmethods{
%Please describe your materials and methods here. This can be more than one paragraph, and may contain subsections and equations as required. Authors should include a statement in the methods section describing how readers will be able to access the data in the paper. 
% }
  Materials and methods can be found in SI Appendix.
  
  \subsection*{Code Availability}
  The source code of the model can be found in Ref.~\citenum{structcol}.
}

\showmatmethods{} % Display the Materials and Methods section
   
\acknow{We thank Rupa Darji, Keith Task, Jerome Fung, Melissa Franklin,
  Mark Schroeder, Bernhard von Vacano, and Rupert Konradi for helpful
  discussions. We also thank Jin-Gyu Park for providing the polystyrene
  nanoparticles used in this work. This work is supported by BASF
  Corporation and the BASF Northeast Research Alliance; by the Harvard
  MRSEC under National Science Foundation (NSF) grant no.\ DMR-2011754;
  and by the NSF Graduate Research Fellowship under grant no.\
  DGE-1144152. It was performed in part at the Harvard Center for
  Nanoscale Systems, supported by NSF grant no.\ 1541959. We thank
  Jeremiah Trimble and Kate Eldridge for assistance with the bird
  feathers and specimen, which were borrowed from the Ornithology
  Department of the Museum of Comparative Zoology, Harvard University.}

\showacknow{} % Display the acknowledgments section

% Bibliography

\end{document}